\documentclass[preprint,aps,superscriptaddress,nofootinbib,tightenlines,floatfix]{revtex4}

\usepackage{graphicx}
\usepackage{bm}
\usepackage{comment}
\usepackage{color}

\def\si{^1 \hskip -0.03in S _0}
\def\siii{^3 \hskip -0.025in S _1}
\def\diii{^3 \hskip -0.045in D _1}

\newcount\hour \newcount\hourminute \newcount\minute 
\hour=\time \divide \hour by 60
\hourminute=\hour \multiply \hourminute by 60
\minute=\time \advance \minute by -\hourminute
\newcommand{\mydate}{\ \today \ - \number\hour :\number\minute}

\begin{document}

\preprint{MIT-CTP 4478, NT@UW-13-22}

\title{Nuclear $\sigma$-terms and Scalar-Isoscalar WIMP-Nucleus
  Interactions from Lattice QCD}

\author{S.R.~Beane} \affiliation{Helmholtz-Institut f\"ur Strahlen-
  und Kernphysik (Theorie), Universit\"at Bonn, D-53115 Bonn, Germany}

\author{S.D.~Cohen} \affiliation{Department of Physics, University of
  Washington, Box 351560, Seattle, WA 98195, USA}

\author{W.~Detmold} \affiliation{ Center for Theoretical Physics,
  Massachusetts Institute of Technology, Cambridge, MA 02139, USA}

\author{H.-W.~Lin} \affiliation{Department of Physics, University of
  Washington, Box 351560, Seattle, WA 98195, USA}

\author{M.J.~Savage} \affiliation{Helmholtz-Institut f\"ur Strahlen-
  und Kernphysik (Theorie), Universit\"at Bonn, D-53115 Bonn,
  Germany}\affiliation{Department of Physics, University of
  Washington, Box 351560, Seattle, WA 98195, USA}

\date{\mydate}

\begin{abstract}
  \noindent
  It has been argued that the leading scalar-isoscalar WIMP-nucleus
  interactions receive parametrically enhanced contributions in the
  context of nuclear effective field theories
  \cite{Prezeau:2003sv}. These contributions arise from meson-exchange
  currents (MECs) and potentially modify the impulse approximation
  estimates of these interactions by 10--60\%. We point out that these
  MECs also contribute to the quark mass dependence of nuclear binding
  energies, that is, nuclear $\sigma$-terms.  In this work, we use
  recent lattice QCD calculations of the binding energies of the
  deuteron, $^3$He and $^4$He at pion masses near 500 MeV and 800 MeV,
  combined with the experimentally determined binding energies at the
  physical point, to provide approximate determinations of the
  $\sigma$-terms for these light nuclei. For each nucleus, we find
  that the deviation of the corresponding nuclear $\sigma$-term from
  the single-nucleon estimate is at the few percent level, in conflict
  with the conjectured enhancement. As a consequence, lattice QCD
  calculations currently indicate that the cross sections for
  scalar-isoscalar WIMP-nucleus interactions arising from fundamental
  WIMP interactions with quarks do not suffer from significant
  uncertainties due to enhanced meson-exchange currents.
\end{abstract}
\maketitle

\section{Introduction}
\label{sec:intro}

\noindent
Nuclei play a central role in laboratory searches for weakly
interacting massive particles (WIMPs), prime candidates for dark
matter, which occur naturally in theoretical frameworks beyond the
Standard Model, such as supersymmetry. WIMPs passing through the local
environment will scatter from nuclei in low-momentum transfer
processes given their expected velocity distribution in our solar
system ($\langle v \rangle\sim 10^{-3} c$)~\cite{Bertone:2004pz},
providing the signature of an isolated recoiling nucleus in a
detector.  Interpretations of candidate WIMP-nucleus events, along
with model predictions for WIMP interaction rates within detectors,
depend upon theoretically challenging calculations of the relevant
nuclear matrix elements.  Typically, the Standard Model fields
involved in the WIMP-nucleus interactions are matched onto operators
in low-energy effective field theories (EFTs) with well-defined
power-counting schemes that describe the dynamics of mesons and
baryons, including chiral perturbation theory
($\chi$PT)~\cite{Weinberg:1978kz,Gasser:1983yg,Gasser:1984gg},
heavy-baryon chiral perturbation theory
(HB$\chi$PT)~\cite{Jenkins:1990jv,Jenkins:1991ne,Bernard:1992qa} and
nuclear chiral
EFTs~\cite{Weinberg:1990rz,Weinberg:1991um,Ordonez:1992xp,vanKolck:1994yi,Kaplan:1996xu,Kaplan:1998tg,Kaplan:1998we,Epelbaum:1998ka,Epelbaum:1999dj,Beane:2008bt}.
Nuclear matrix elements of these hadronic-level operators are then
calculated with nuclear many-body techniques to provide the
WIMP-nucleus interactions and their associated couplings.  At present,
it remains challenging to perform both stages of the calculation.  The
first stage requires solving QCD, which is beginning to be
accomplished with the numerical technique of lattice QCD at the
physical light-quark masses. The second stage requires forming matrix
elements of the hadronic-level operators in a nonrelativistic
interacting quantum many-body
system~\cite{Prezeau:2003sv,Fitzpatrick:2012ix,Cirigliano:2012pq,Fitzpatrick:2012ib,Menendez:2012tm,Klos:2013rwa}.
At some point in the future, lattice QCD calculations will be able to
determine such nuclear matrix elements in light nuclei by a direct
evaluation, obviating the need for either matching step. However, for
the foreseeable future, the less direct approach in which lattice QCD
is used to constrain couplings in nuclear chiral EFTs is necessary.

An important feature of nuclei is that, to a large extent, they behave
as a collection of nonrelativistic nucleons dominated by two-body
interactions, with the three-body and higher interactions strongly
suppressed.  Such a hierarchy of forces is understood within the
frameworks of low-energy chiral nuclear forces.  As such, many
low-energy nuclear observables are dominated by the contributions from
individual nucleons (the impulse approximation), which would yield the
naive estimate
\begin{eqnarray}
  g_{Z,N} & = & g_p Z\ +\ g_n N
  \ \ ,
  \label{eq:naive}
\end{eqnarray}
for a WIMP-nucleus coupling, where $g_{p,n}$ are the appropriate WIMP
couplings to protons and neutrons, respectively, and $N$ and $Z$ are
the neutron and proton numbers.  Typically, nuclear interactions (for
instance, meson-exchange currents (MECs)) are expected to correct the
impulse approximation result at the few-percent level.  However, it
has been argued that scalar WIMP interactions with nuclei might
violate this hierarchy due to a parametric enhancement of MECs
involving the up and down quarks in the low-energy chiral EFT,
providing a correction to the impulse approximation, and hence a
correction to Eq.~(\ref{eq:naive}), at the 10--60\%
level~\cite{Prezeau:2003sv}.  It is important to note that such
contributions could have a dependence upon $Z$ and $N$ that is
substantially different from that given in Eq.~(\ref{eq:naive}), and
thereby could provide significantly more freedom in the relative
contributions for different nuclei.  The conjectured enhancement of
isoscalar MECs would imply that our knowledge of WIMP-nucleus cross
sections is significantly less certain than often assumed and may
differ substantially among different nuclear targets.  Currently, the
most stringent constraints on spin-independent WIMP-nucleon
interactions come from the XENON collaboration~\cite{Aprile:2010bt}
(see Ref.~\cite{Beltrame:2013bba}) using the impulse approximation
(see also limits from CDMS~\cite{Akerib:2005kh},
CRESST-II~\cite{Angloher:2008jj}, ZEPLIN-III~\cite{Akimov:2011tj},
EDELWEISS-II~\cite{Armengaud:2012pfa}, WARP~\cite{Benetti:2007cd},
TEXONO \cite{Li:2013fla}, and CDEX \cite{Zhao:2013xsf}).  For
instance, the XENON collaboration finds that the WIMP-nucleon cross
section is less than $2\times 10^{-45}~{\rm cm}^2$ for a WIMP with a
mass of $M_\chi=55~{\rm GeV}$ at the $90\%$ confidence
level~\cite{Beltrame:2013bba}.  However, the
DAMA~\cite{Bernabei:2010mq} and CoGeNT~\cite{Aalseth:2010vx}
experiments claim evidence of a light WIMP, in the $\sim 10$~GeV
range, including statistically significant annual modulation (as an
example of analyses of the consistency of the experimental results,
see Ref.~\cite{Fox:2011px}).  Given the different nuclei used as
targets in these experiments, it is possible that an enhanced nuclear
contribution with different dependence on $N$ and $Z$ could decrease
the tension amongst these different results \cite{Prezeau:2003sv}.  In
this work, we use recent lattice QCD calculations of the mass of the
deuteron, $^3$He and $^4$He at pion masses of $m_\pi\sim 510~{\rm
  MeV}$ ~\cite{Yamazaki:2012hi,Yamazaki:2012fn} and $\sim 806~{\rm
  MeV}$~\cite{Beane:2012vq}, combined with the known masses in nature,
to provide constraints on the matrix elements of the light-quark
scalar-isoscalar interaction in these light nuclei directly from QCD
via their nuclear 
$\sigma$-terms~\footnote{
The lattice QCD based
  approach of Refs.~\cite{Aoki:2012tk} does not find a
  bound deuteron at the SU(3)-symmetric point, in seeming disagreement
  with Refs.~\cite{Beane:2012vq,Yamazaki:2012hi,Yamazaki:2012fn}. 
  Solving the Schrodinger equation with the energy-dependent ``potentials'' defined in the QCD modeling method recovers the scattering phase-shifts of QCD
  with the same level of rigor as the L\"uscher method only at the energy-eigenvalues of the lattice calculation that gave rise to the potentials.
  Away from those energy eigenvalues, the HALQCD modeling method produces results that are not those of QCD. 
  However, it should be stressed that at the energy eigenvalues of the lattice  calculation, the HALQCD modeling 
  method does provide a phase-shift and binding energy at the same level of rigor as that obtained with L\"uscher's method.
  HALQCD trades the energy dependence of the potentials for nonlocality, then approximates the nonlocality by locality, and has performed calculations showing that the energy-dependence is weak compared with statistical uncertainties in their calculations. Further, they fit these now local potentials to a finite number of Gaussian
  profiles, along with Yukawa interactions that are softened at short distance.
While one would expect that applying this method to the deuteron binding energy calculated in sufficiently large lattice volumes 
 would result in a deuteron binding energy that is close to an energy eigenvalue, they have not presented the eigenvalues.  
Since solving the discrepancy is not the subject of this work, 
only the results obtained with L\"uscher's method are used.
 }.  
We find no evidence for enhanced
deviations from the single-nucleon estimate for these interactions,
instead finding them to be at the few-percent level, consistent with
nuclear effects in other observables.

\section{Nuclear $\mathbf{\sigma}$-Terms and Scalar WIMP-Nucleus
  Interactions}
\label{sec:SigmaNucleus}
\noindent
Over a significant range of WIMP masses, the strength of their
interactions with hadrons is tightly constrained by experimental
searches.  However, in the absence of definitive observations in
multiple processes, the structure of the interaction between WIMPs and
Standard Model fields remains unknown.  Restricting our discussion to
scalar interactions between WIMPs and dimension-three quark
bilinears\footnote{In many cases, these are the most relevant
  operators, but there are dark matter candidates for which 
  scalar-isoscalar gluonic operators are important, such as
  technibaryon dark matter
  \cite{Nussinov:1992he,Chivukula:1992pn,Bagnasco:1993st}.}, the
leading-order (LO) interactions of spin-${1\over 2}$ WIMPs, $\chi$, in
an expansion in derivatives, can be written as
\begin{eqnarray} {\cal L} & = & G_F \ \overline{\chi}\chi\ \sum_q \
  a_S^{(q)} \ \overline{q} q \ =\ G_F \ \overline{\chi}\chi\
  \overline{q} a_S q
  \nonumber\\
  & = & {G_F\over 2} \ \overline{\chi}\chi\ \left[
    (a_S^{(u)}+a_S^{(d)}) \overline{q}q \ +\ (a_S^{(u)}-a_S^{(d)})
    \overline{q}\tau^3 q \ +\ 2 \ a_S^{(s)} \overline{s}s\ +\ ...\ \right]
  \ \ \ ,
  \label{eq:scalarquarklevel}
\end{eqnarray}
where $ \overline{q} q = \overline{u}u+\overline{d}d$, and where $G_F$
is Fermi's weak coupling constant --- here used to normalize the
(scale-dependent) couplings of the WIMP(s) to the quarks, $a_S^{(q)}$,
with respect to a weak-scale interaction strength.  For two light
flavors $a_S = {\rm diag}(a_S^{(u)},a_S^{(d)})$ and for three light
flavors $a_S = {\rm diag}(a_S^{(u)},a_S^{(d)},a_S^{(s)})$.  The
generalization to WIMPs with arbitrary spin, but with scalar
interactions with the quarks, is obvious.  The part of the operator
involving Standard Model fields transforms in the same way as the
light-quark mass terms in the QCD Lagrange density, and consequently,
its leading hadronic matrix elements are known.  Based upon the
spontaneously broken $SU(2)_L\otimes SU(2)_R$ chiral symmetry of QCD,
the interaction Lagrange density in Eq.~(\ref{eq:scalarquarklevel})
matches onto
\begin{eqnarray} {\cal L}
  & \rightarrow & 
  G_F  \ 
  \overline{\chi}\chi\ 
  \left(\ 
    {1\over 4} \langle 0|\overline{q}q |0\rangle\ 
    {\rm Tr}\left[ a_S\Sigma^\dagger + a_S^\dagger\Sigma \right]
    \ +\ 
    {1\over 4}\langle N|\overline{q}q |N\rangle N^\dagger N 
    {\rm Tr}\left[ a_S\Sigma^\dagger + a_S^\dagger\Sigma \right]
  \right.
  \nonumber\\
  &&
  \left.
    \qquad\qquad
    \ -\ 
    {1\over 4}
    \langle N|\overline{q}\tau^3 q |N\rangle 
    \left(
      N^\dagger N 
      {\rm Tr}\left[ a_S\Sigma^\dagger + a_S^\dagger\Sigma \right]
      \ -\ 4 N^\dagger  a_{S,\xi} N 
    \right)
    \ +\ ...
  \right)
  \ \ \ 
  \label{eq:scalarhadronlevel}
\end{eqnarray}
at the chiral symmetry breaking scale $\Lambda_\chi$, which describes
the single-hadron matrix elements and the associated interactions at
LO in the chiral expansion.  $\Sigma$ is the exponentiated pion field,
and $N$ is the nucleon field,
\begin{eqnarray}
  \Sigma & = & \exp\left({2i\over f_\pi} M\right)
  \ \ ,\ \ 
  M\ =\ 
  \left(
    \begin{array}{cc}
      \pi^0/\sqrt{2} & \pi^+ \\
      \pi^- & -\pi^0/\sqrt{2}
    \end{array}
  \right)
  \ \ ,\ \ 
  N\ =\ 
  \left(
    \begin{array}{c}
      p\\
      n
    \end{array}
  \right)
  \ \ \ ,
  \label{eq:mesons}
\end{eqnarray}
$f_\pi=132$ MeV is the pion decay constant, $a_{S,\xi} = {1\over
  2}\left(\xi^\dagger a_S \xi^\dagger + \xi a_S^\dagger \xi\right)$
with $\xi=\sqrt{\Sigma}$, and the ellipsis denotes higher-order
interactions including those involving more than one nucleon.
Expanding Eq.~(\ref{eq:scalarhadronlevel}) in the number of pion
fields (neglecting the shift in the WIMP mass induced by the chiral
condensate), the LO contributions to the interactions are
\begin{eqnarray} {\cal L}
  & \rightarrow & 
  G_F  \ 
  \overline{\chi}\chi\ 
  \Bigg(
  -{(a_S^{(u)}+a_S^{(d)})\over f_\pi^2} \langle 0|\overline{q}q |0\rangle\ 
  \ \left({1\over 2}(\pi^0)^2 + \pi^+\pi^- \right)
  \ +\ 
  {1\over 2}(a_S^{(u)}+a_S^{(d)})\langle N|\overline{q}q |N\rangle
  N^\dagger N 
  \nonumber\\
  &&
  \qquad\qquad
  \ +\ 
  {1\over 2} (a_S^{(u)}-a_S^{(d)})
  \langle N|\overline{q}\tau^3 q |N\rangle 
  N^\dagger  \tau^3 N 
  \ +\ ...
  \Bigg)
  \ \ \ .
  \label{eq:scalarhadronlevelexp}
\end{eqnarray}
Matching onto the multi-nucleon interactions is complicated by the
fact that contributions from pion-exchange interactions and from local
four-nucleon operators are of the same order in the chiral expansion,
and the coefficients of the latter are not directly related to
multi-nucleon matrix elements at any order in the chiral expansion.
For instance, the four-nucleon operators involving one insertion of
the light-quark mass matrix are of the
form~\cite{Kaplan:1996xu,Kaplan:1998tg,Kaplan:1998we}
\begin{eqnarray} {\cal L}^{N^4, m_q} & = & D_{S,1} \left(N^\dagger
    N\right)^2 {\rm Tr}\left[ m_q\Sigma^\dagger + m_q^\dagger\Sigma
  \right] \ +\ D_{S,2} N^\dagger N N^\dagger m_{q,\xi+} N
  \nonumber\\
  & + & D_{T,1} \left(N^\dagger \sigma^a N\right)^2 {\rm Tr}\left[
    m_q\Sigma^\dagger + m_q^\dagger\Sigma \right] \ +\ D_{T,2}
  N^\dagger \sigma^a N N^\dagger \sigma^a m_{q,\xi+} N \ \ \
  \label{eq:4Nmass}
\end{eqnarray}
in the low-energy EFT, where $m_{q,\xi+} = {1\over 2}\left(\xi^\dagger
  m_q \xi^\dagger + \xi m_q^\dagger \xi\right)$, and $\sigma^a$ are
the Pauli matrices. Hence WIMP--two-nucleon interactions are of the
form
\begin{eqnarray} {\cal L}^{N^4, \chi} & = & -G_F \overline{\chi}\chi
  \left( D_{S,1} \left(N^\dagger N\right)^2 {\rm Tr}\left[
      a_S\Sigma^\dagger + a_S^\dagger\Sigma \right] \ +\ D_{S,2}
    N^\dagger N N^\dagger a_{S,\xi} N \right.
  \nonumber\\
  && \left.  + D_{T,1} \left(N^\dagger \sigma^a N\right)^2 {\rm
      Tr}\left[ a_S\Sigma^\dagger + a_S^\dagger\Sigma \right] \ +\
    D_{T,2} N^\dagger \sigma^a N N^\dagger \sigma^a a_{S,\xi} N
  \right) \ \ \ .
  \label{eq:4Nwimp}
\end{eqnarray}

\begin{figure}[!ht]
  \centering
  \raisebox{3ex}{\includegraphics[width=0.25\textwidth]{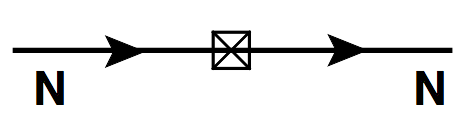}
  }\qquad\qquad
  \includegraphics[width=0.25\textwidth]{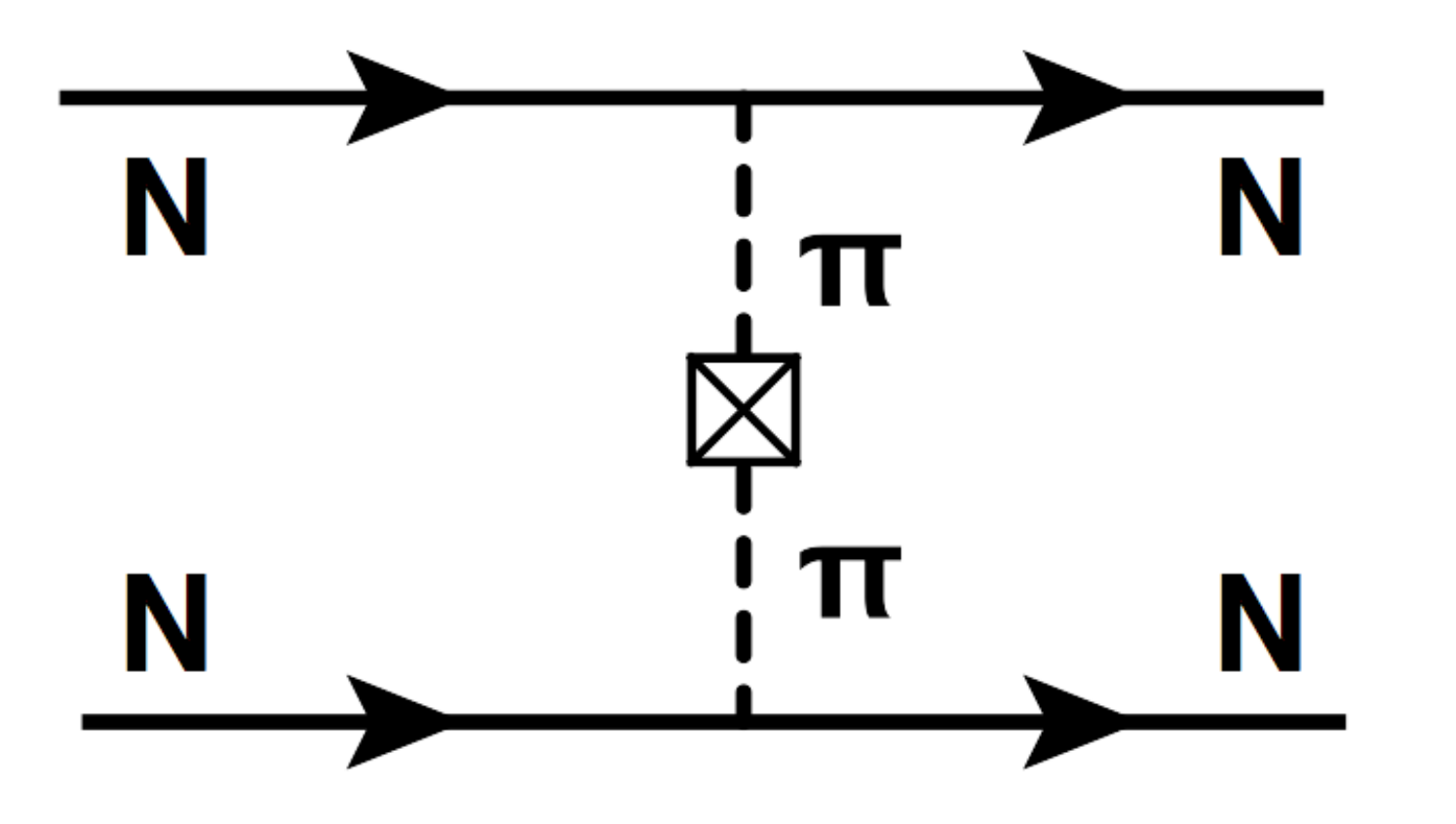}\qquad\qquad
  \includegraphics[width=0.25\textwidth]{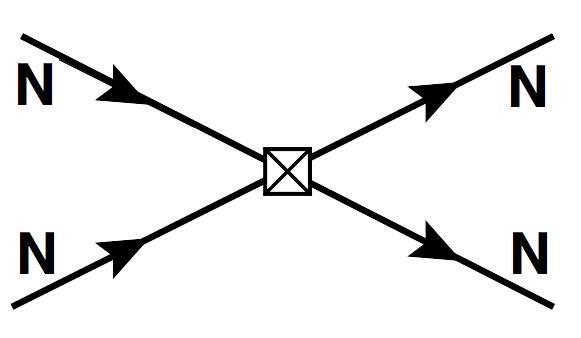}
  \caption{ Some of the diagrams contributing to nuclear
    $\sigma$-terms. The left panel shows the leading order
    contribution to the single-nucleon $\sigma$-term in $\chi$PT.  The
    middle (pion-exchange) and right ($``D_2$-terms'' contributions
    from Eq.~(\protect\ref{eq:4Nwimp})) panels show contributions to
    nuclear $\sigma$-terms at next-to-leading order in KSW power
    counting~\cite{Kaplan:1996xu,Kaplan:1998tg,Kaplan:1998we}.  The
    crossed box corresponds to an insertion of the light-quark mass
    matrix.  }
  \label{fig:diagramsA}
\end{figure}

The importance of the various contributions to the scalar-isoscalar
matrix elements can be estimated using power counting arguments.  The
second and third terms in Eq.~(\ref{eq:scalarhadronlevelexp}) provide
the leading (order $Q^0$, where $Q$ denotes the small ratio of scales
in the effective theory) scalar interactions between the WIMP and the
nucleon that generate the impulse approximation for WIMP-nucleus
interactions (see Fig.~\ref{fig:diagramsA} (left)).  In a nucleus, the
first term in Eq.~(\ref{eq:scalarhadronlevelexp}) gives rise to a MEC
between two nucleons, as shown in Fig.~\ref{fig:diagramsA} (middle),
that naively contributes at order $1/Q^2$ in the chiral expansion due
to the non-derivative interaction of the pions, which is two orders
lower than the contribution from the impulse approximation.  This term
is the origin of the enhancement suggested in
Ref.~\cite{Prezeau:2003sv}.  The isoscalar interactions with the
strange and heavier quarks do not contribute to the non-derivative
interaction with pions and, as such, are not expected to be enhanced
in WIMP-nucleus interactions. To determine the WIMP-nucleus
interactions quantitatively, nuclear matrix elements of these
operators need to be calculated.

Ideally, one would simply determine the matrix element of the Lagrange
density in Eq.~(\ref{eq:scalarquarklevel}) in the ground state of a
given nucleus, at the relevant momentum transfer, without performing
the intermediate matchings in Eq.~(\ref{eq:scalarhadronlevel}) and in
Eq.~(\ref{eq:scalarhadronlevelexp}).  This would sum the contributions
from the hadronic EFT to all orders in perturbation theory, and
provide the necessary matrix elements directly from QCD.  While such
formidable calculations cannot currently be accomplished, the forward
matrix element of the scalar-isoscalar operator can be determined in
light nuclei, albeit with significant uncertainties, by combining
recent lattice QCD calculations of the binding energies with the
corresponding experimental values.  The mass of the ground state of a
nucleus with $Z$ protons and $N$ neutrons, denoted by $|Z,N
({\rm gs})\rangle$, is $E^{({\rm gs})}_{Z,N} = E^{({\rm gs},\chi)}_{Z,N} +
\sigma_{Z,N}$, where
\begin{eqnarray}
  \sigma_{Z,N} & = & 
  \langle Z,N ({\rm gs})| 
  \ m_u \overline{u}u + m_d \overline{d}d 
  \ | Z,N ({\rm gs})\rangle
  \ \ \ 
  \label{eg:ZNsigdef}
\end{eqnarray}
is the nuclear $\sigma$-term, and $ E^{({\rm gs},\chi)}_{Z,N}$ is the
energy of the nuclear ground state in the limit of massless up- and
down-quarks (assuming that the nucleus is bound in this limit).  With
isospin symmetry, $m_u=m_d=\overline{m}$, the nuclear $\sigma$-term
becomes
\begin{eqnarray}
  \sigma_{Z,N} & = & 
  \overline{m}
  \langle Z,N ({\rm gs})| 
  \ \overline{u}u + \overline{d}d 
  \ | Z,N ({\rm gs})\rangle
  \ =\ 
  \overline{m}{d\over d\overline{m}} E^{({\rm gs})}_{Z,N}
  \nonumber\\
  & = & 
  \left[\ 1\ +\ {\cal O}\left(m_\pi^2\right)\ \right]
  {m_\pi\over 2}{d\over d m_\pi}  E^{({\rm gs})}_{Z,N}
  \ \ \ ,
  \label{eg:ZNsigdefisospin}
\end{eqnarray}
where we have used the leading contribution to the
Gell-Mann--Oakes--Renner (GMOR)
relation~\cite{GellMann:1968rz,Gasser:1983yg},
\begin{eqnarray}
  -2\overline{m} 
  \langle 0| 
  \ \overline{u}u + \overline{d}d 
  \ | 0\rangle
  \ =\ 
  m_\pi^2 f_\pi^2
  \left[\ 1\ +\ {\cal O}\left(m_\pi^2\right)\ \right]
  \ \ \ ,
  \label{eg:gmor}
\end{eqnarray}
to relate the quark and pion masses.  The relation between the pion
mass and the average light-quark mass has been precisely determined
with lattice QCD~\cite{Lin:2006cf,Aoki:2010dy}.  The linear relation
between $m_\pi^2$ and $\overline{m}$ is found to hold to better than
$10\%$ over a large range of pion masses, even for the heavy pion
masses that we consider~\cite{Lin:2006cf,Aoki:2010dy}.  We use this
linear relationship in constructing nucleon and nuclear $\sigma$
terms, $\sigma_{Z,N} = {m_\pi\over 2}{d\over d m_\pi} E^{({\rm
    gs})}_{Z,N}$, and assign a conservative $10\%$ uncertainty in
order to account for the nonlinearity in the GMOR relation (note that
this uncertainty will later cancel when we take the ratio of
$\sigma$-terms below).

Writing the mass of the nucleus as $E^{({\rm gs})}_{Z,N} = A M_N -
B_{Z,N}$, where $A=Z+N$, $M_N$ is the isospin-averaged nucleon mass,
and $B_{Z,N}$ is the total binding energy of the nucleus, the
(two-flavor) nuclear $\sigma$-term can be written as
\begin{eqnarray}
  \sigma_{Z,N} & = & 
  A \sigma_{N} + \sigma_{B_{Z,N}}
  \ = \ 
  A \sigma_{N} - {m_\pi\over 2}{d\over d m_\pi} B_{Z,N}
  \ \ \ ,
  \label{eg:ZNsigdefisospinbinding}
\end{eqnarray}
where
\begin{eqnarray}
  \sigma_{N} & = & 
  \overline{m}
  \langle N| 
  \ \overline{u}u + \overline{d}d 
  \ | N\rangle
  \ =\ 
  \overline{m}{d\over d\overline{m}} M_{N}
  \ =\ 
  {m_\pi\over 2}{d\over d m_\pi}  M_{N}
  \label{eg:Nsigdefisospin}
\end{eqnarray}
is the nucleon $\sigma$-term and $|N\rangle$ is the single-nucleon
state.  The first term in Eq.~(\ref{eg:ZNsigdefisospinbinding}) is the
noninteracting single-nucleon contribution to the nuclear
$\sigma$-term, while the second term corresponds to the corrections
due to interactions between the nucleons, including the possibly
enhanced contributions from MECs.  It is useful to define the ratio
\begin{eqnarray}
  \delta\sigma_{Z,N} & = & 
  -
  {1 \over A \sigma_{N}}
  {m_\pi\over 2}{d\over d m_\pi} B_{Z,N}
  \ \ \ 
  \label{eg:ZNsigrat}
\end{eqnarray}
to quantify the deviations from the impulse approximation.  In
addition to representing deviations of nuclear $\sigma$-terms from the
impulse approximation, this quantity also describes the deviation of
the scalar-isoscalar WIMP-nucleus scattering matrix element from the
impulse approximation at zero momentum transfer,
\begin{eqnarray}
  \delta\sigma_{Z,N} & = & 
  {
    \langle Z,N ({\rm gs})|\ \overline{u}u+\overline{d}d | Z,N ({\rm gs})\rangle
    \over
    A \ \langle N|\ \overline{u}u+\overline{d}d | N\rangle
  }
  \ -\ 1
  \ \ \ .
  \label{eq:explicitratio}
\end{eqnarray}

\section{Light Nuclei from Lattice QCD and their
  $\mathbf{\sigma}$-Terms}
\label{sec:LQCDresults}
\noindent
Lattice QCD has evolved to the stage where the binding energies of the
lightest nuclei and hypernuclei have been determined at a small number
of relatively heavy pion masses in the limit of isospin symmetry.
Further, the mass of the nucleon has been explored extensively over a
large range of light-quark masses, with calculations now being
performed at the physical value of the pion mass.  These sets of
calculations, along with the experimental values of the masses of the
light nuclei, are sufficient to arrive at a first QCD determination of
the nuclear $\sigma$-terms for these nuclei at a small number of pion
masses. This work provides an estimate of the modifications to the
impulse approximation for scalar-isoscalar WIMP-nucleus interactions
in light nuclei. In particular, these results can be used to explore
the conjectured enhancement of MEC contributions to these
interactions, and to investigate the size of the uncertainties
introduced by the use of the impulse approximation in phenomenological
analyses.  It is important to mention that the EFT description of the
quark-mass dependence of the nuclear forces has been developed in
Refs.~\cite{Beane:2002vs,Epelbaum:2002gb,Beane:2002xf,Beane:2006mx,Soto:2011tb},
and estimates of nuclear $\sigma$ terms have been made in
Refs.~\cite{Flambaum:2007mj,CarrilloSerrano:2012ja,Berengut:2013nh,Epelbaum:2012iu,Epelbaum:2013wla}.
We will make use of EFT below in assessing a systematic uncertainty due to
extrapolation in the light-quark masses.

\begin{figure}[!ht]
  \centering
  \includegraphics[width=0.31\textwidth]{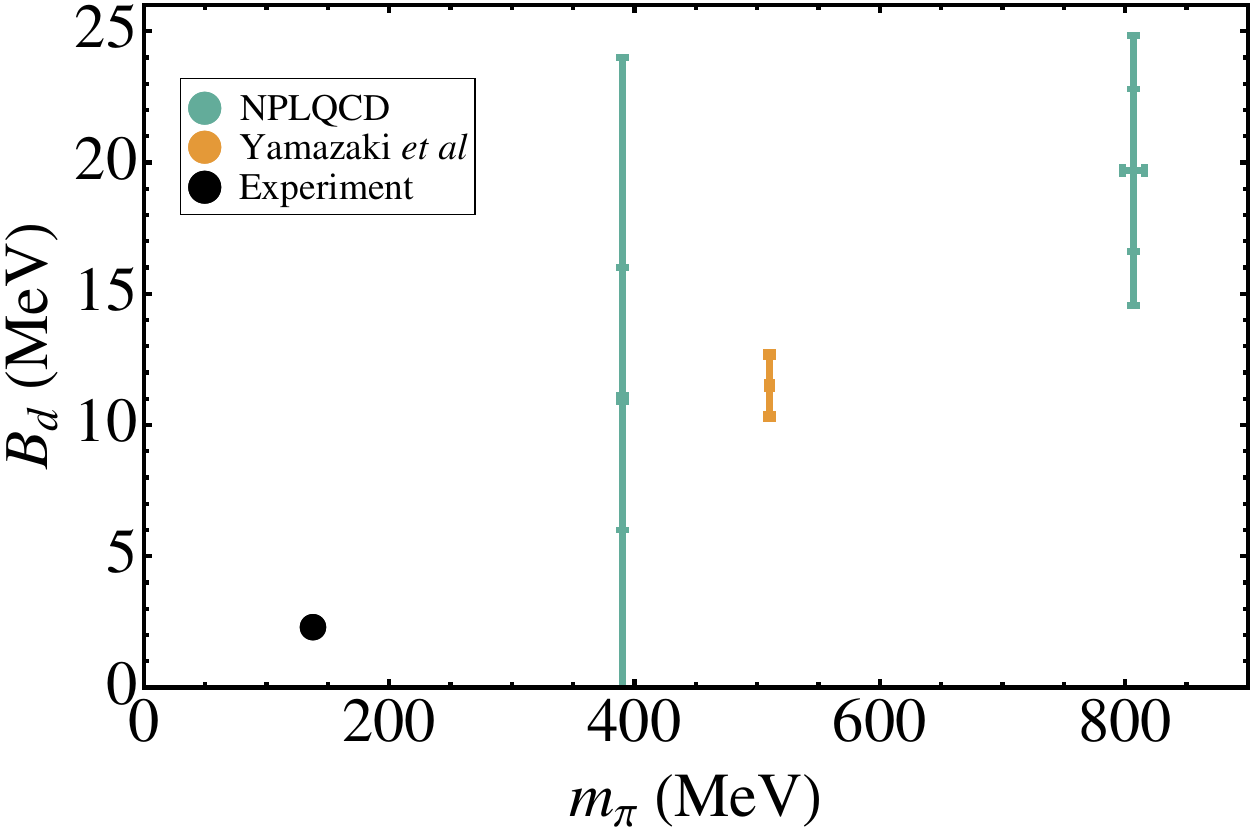}\ \quad
  \includegraphics[width=0.31\textwidth]{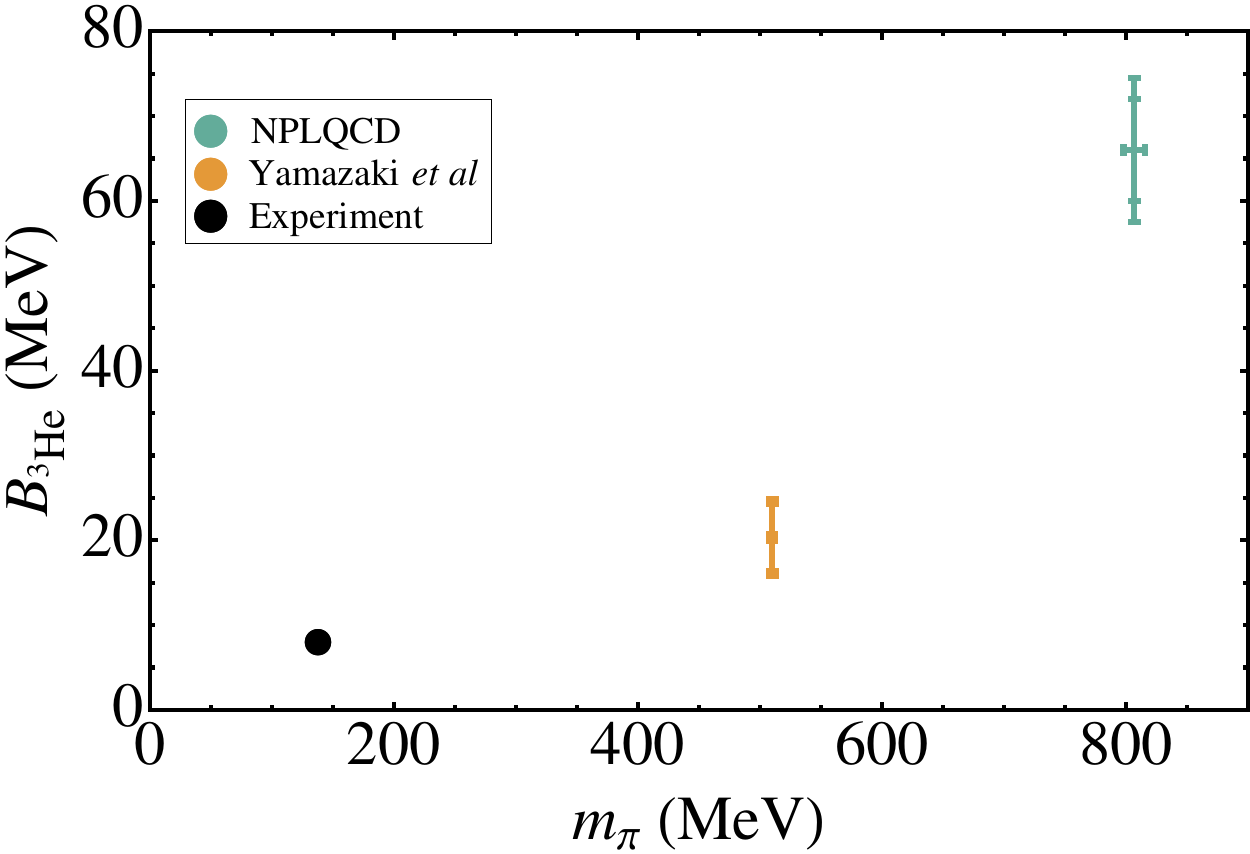}\ \quad
  \includegraphics[width=0.31\textwidth]{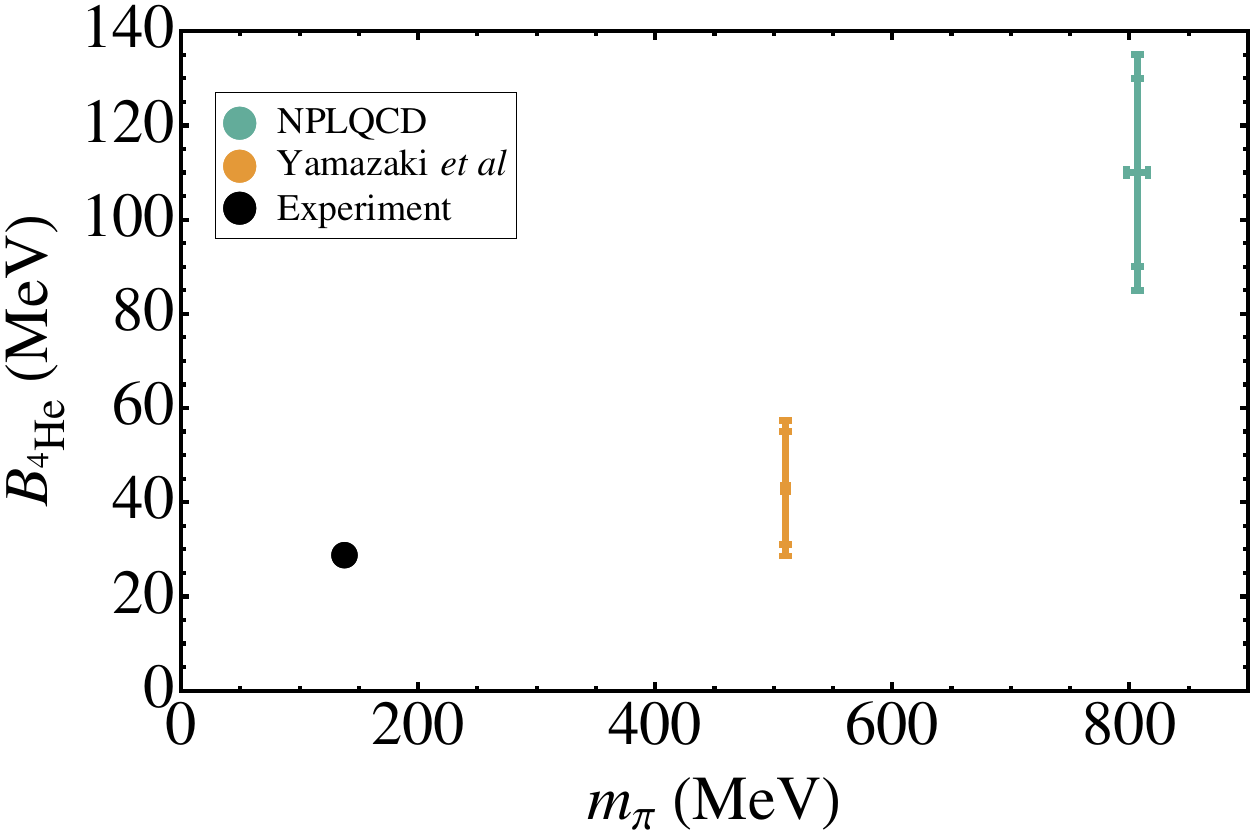}
  \caption{The deuteron (left panel), $^3$He (middle panel) and $^4$He
    (right panel) binding energies from $n_f=2+1$ lattice QCD
    calculations, along with the experimental values.  The inner and
    outer uncertainty bars correspond to the statistical and total
    (statistical combined with systematic) uncertainties,
    respectively.}
  \label{fig:Ball}
\end{figure}

\begin{figure}[!ht]
  \centering
  \includegraphics[width=0.55\textwidth]{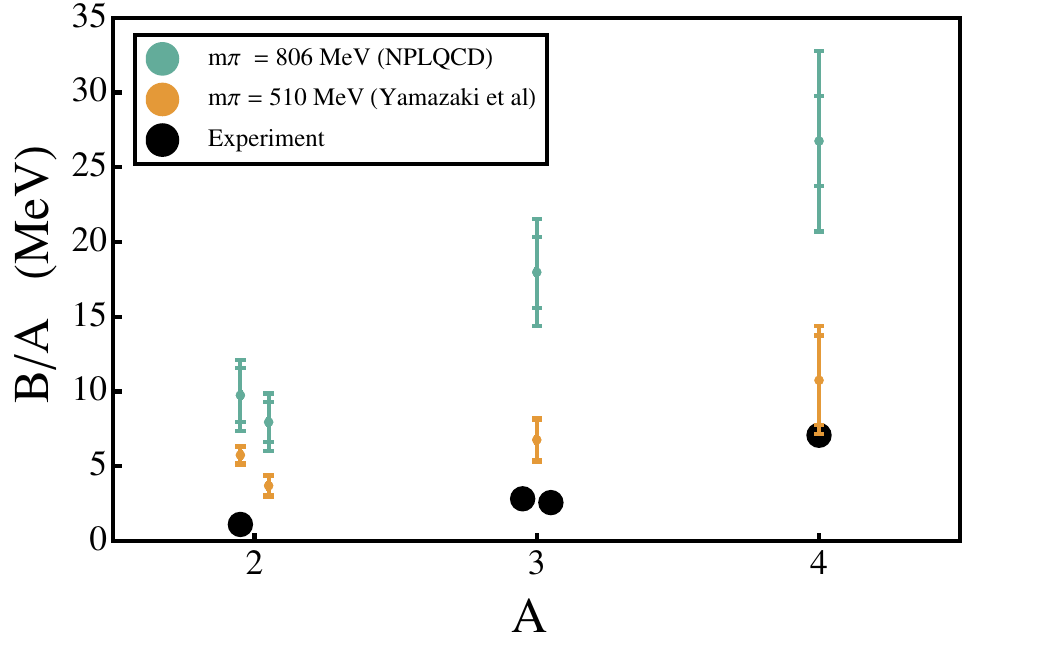}
  \caption{ The binding energy per nucleon calculated with lattice
    QCD~\cite{Yamazaki:2012hi,Beane:2012vq,Yamazaki:2012fn}, in the
    limit of isospin symmetry and the absence of electromagnetic
    interactions, along with their experimental values, as given in
    Table~\ref{tab:Masses}.  The two sets of calculations shown with
    $A=2$ correspond to the deuteron (left) and the dineutron (right),
    the latter of which is unbound for the physical quark masses.  }
  \label{fig:BoverA}
\end{figure}

The binding energies of the deuteron, $^3$He and $^4$He at pion masses
of $m_\pi\sim 390$, $510$ and $806~{\rm MeV}$ calculated with lattice
QCD~\cite{Beane:2009py,Beane:2011iw,Yamazaki:2012hi,Beane:2012vq,Yamazaki:2012fn}
are presented in Table~\ref{tab:Masses}, along with their values at
the physical point, and are shown in Fig.~\ref{fig:Ball}. The binding
energies per nucleon are shown in Fig.~\ref{fig:BoverA}.
\begin{table}
  \begin{center}
    \begin{minipage}[!ht]{16.5 cm}
      \caption{ The binding energies of the deuteron, $^3$He and
        $^4$He at pion masses of $m_\pi\sim 390~{\rm
          MeV}$~\cite{Beane:2009py,Beane:2011iw}, $\sim 510~{\rm
          MeV}$~\cite{Yamazaki:2012hi,Yamazaki:2012fn}, and $\sim
        806~{\rm MeV}$~\cite{Beane:2012vq} calculated with lattice
        QCD, along with their values at the physical point (the
        uncertainties in the experimental measurements are
        negligible).  }
      \label{tab:Masses}
    \end{minipage}
    \setlength{\tabcolsep}{1em} \resizebox{0.9 \linewidth}{!}{%
      \begin{tabular}{c|ccc}
        \hline
        $m_\pi~({\rm MeV})$
        & $B_d$ (MeV) 
        & $B_{^3 \rm He}$ (MeV) 
        & $B_{^4 \rm He}$ (MeV)   \\
        \hline
        138.0 (expt, isospin-averaged)
        & 2.22 
        & 7.718 
        & 28.3  \\
        $390.3(2.6)$~\cite{Beane:2009py,Beane:2011iw}
        & 11(5)(12)
        & -
        & - \\
        $510(2)$~\cite{Yamazaki:2012hi,Yamazaki:2012fn}
        & 11.5(1.1)(0.6)
        & 20.3(4.0)(2.0)
        & 43(12)(8) \\
        $806.7(8.9)$~\cite{Beane:2012vq}
        & 19.7(3.1)(4.1)
        & 66(6)(6)
        & 110(20)(15)\\
        \hline
      \end{tabular}
    }
    \begin{minipage}[t]{16.5 cm}
      \vskip 0.0cm
      \noindent
    \end{minipage}
  \end{center}
\end{table}     
The lattice QCD calculations were performed with clover-improved
discretizations of the quark fields.  The $m_\pi \sim 806$ MeV
calculations were performed with a lattice spacing of $b \sim 0.15$ fm
(determined at this mass) \cite{Beane:2012vq}. The $m_\pi \sim 390$
MeV calculations\footnote{Since the calculations with $m_\pi~\sim
  390~{\rm MeV}$ did not determine the binding energies of the $A=3,4$
  systems, and extracted the binding energy of the deuteron with large
  uncertainties, we do not use these results in our analysis of the
  sigma terms.} used an anisotropic discretization and were performed
with a spatial lattice spacing $b_s \sim 0.12$ fm (when determined by
extrapolation to the physical quark
masses)~\cite{Beane:2009py,Beane:2011iw}. Finally, the $m_\pi \sim
510$ MeV calculations were performed with $b \sim 0.09$
fm~\cite{Yamazaki:2012hi,Yamazaki:2012fn}.
Each set of calculations was performed in multiple lattice volumes to
distinguish continuum states from bound states, but none of the
calculations were extrapolated to the continuum limit, leading to a
small additional uncertainty (the binding energies are expected to
have uncertainties of ${\cal O}(b^2)$) not shown in
Table~\ref{tab:Masses} that we neglect in our analysis.  In each set
of calculations, the strange-quark mass was tuned to be approximately
its physical value.  The small systematic uncertainty in the nuclear
$\sigma$-terms due to the inexact tuning of the strange-quark mass is
also neglected in this analysis.  Further details of these sets of
calculations can be found in the original
references~\cite{Beane:2009py,Beane:2011iw,Yamazaki:2012hi,Beane:2012vq,Yamazaki:2012fn},
and we do not repeat them here.

\begin{figure}[!ht]
  \centering
  \includegraphics[width=0.31\textwidth]{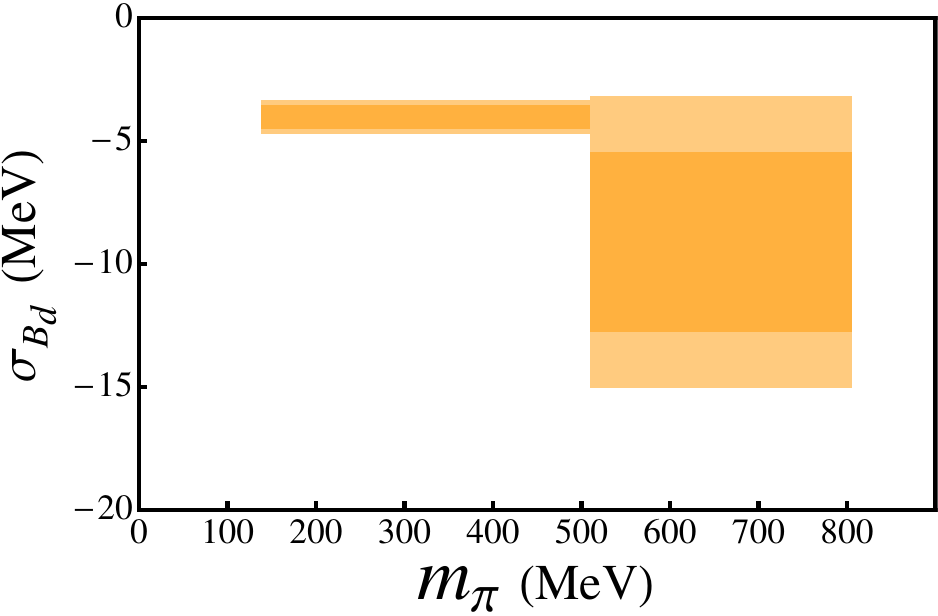}
  \quad
  \includegraphics[width=0.31\textwidth]{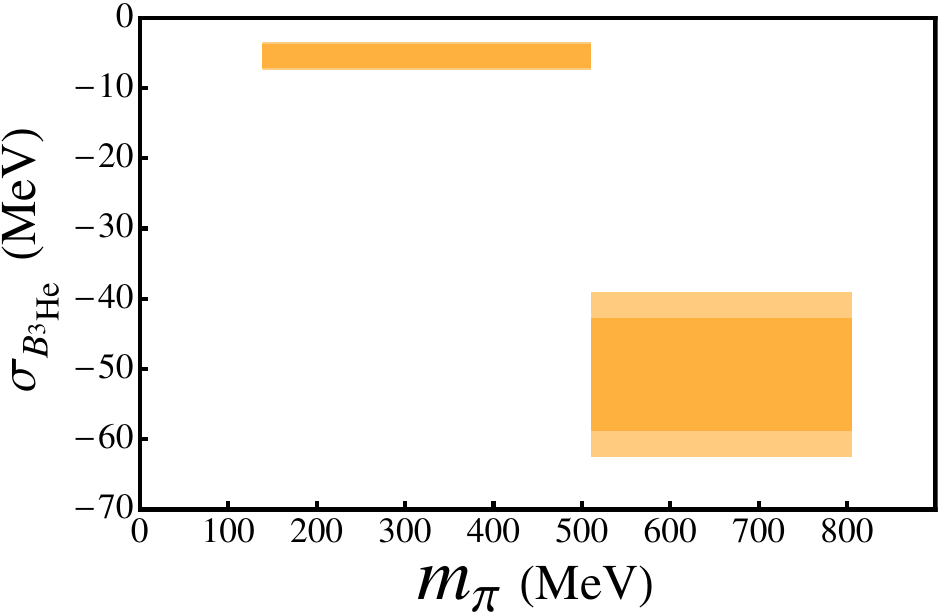}
  \quad
  \includegraphics[width=0.31\textwidth]{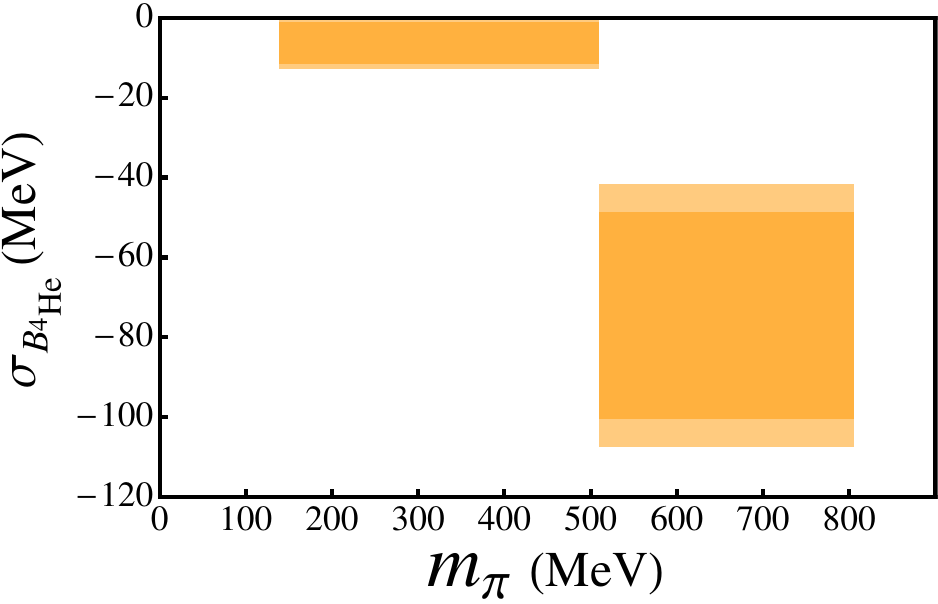}
  \caption{ The nuclear contributions to the deuteron (left panel),
    $^3$He (middle panel) and $^4$He (right panel) $\sigma$-terms from
    nuclear interactions. The inner and outer shaded regions
    correspond to the statistical and total (statistical combined with
    systematic) uncertainties, respectively.}
  \label{fig:SigmaB}
\end{figure}

\begin{figure}[!ht]
  \centering
  \includegraphics[width=0.31\textwidth]{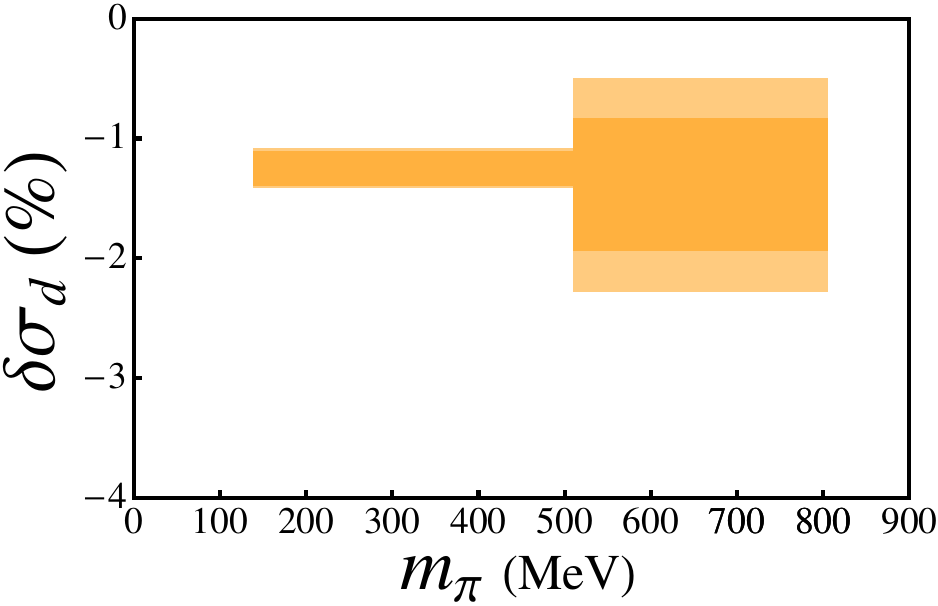}\quad
  \includegraphics[width=0.31\textwidth]{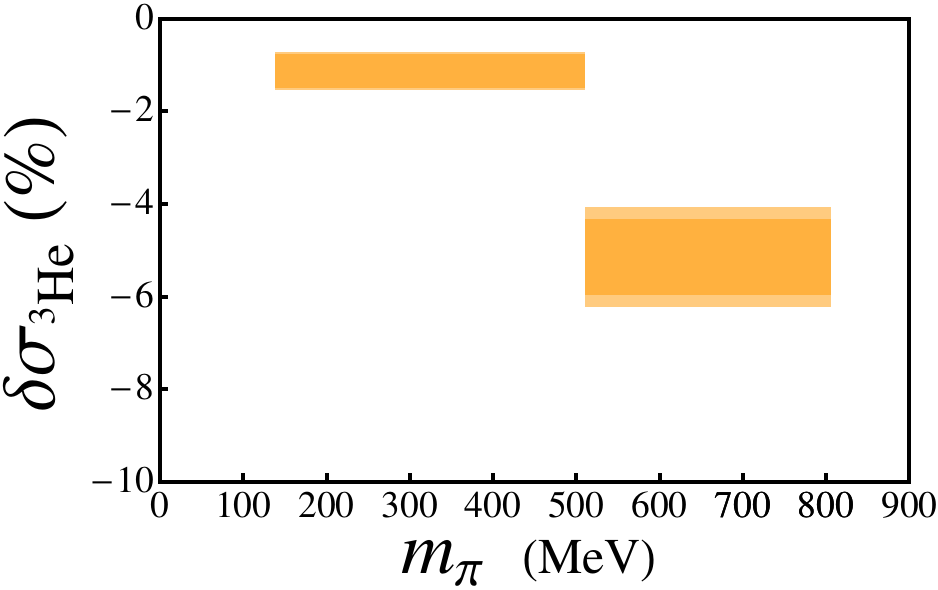}\quad
  \includegraphics[width=0.31\textwidth]{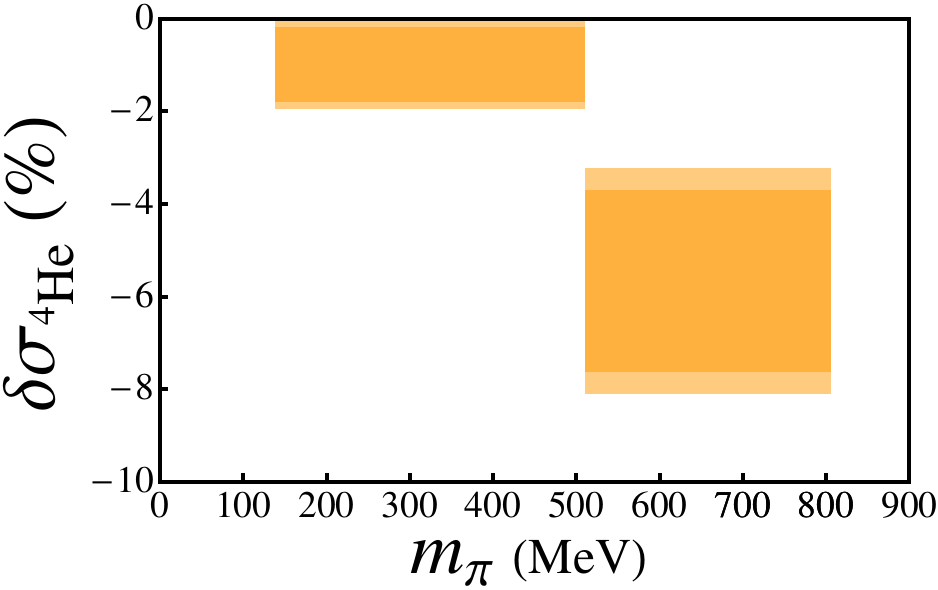}
  \caption{ Percentage modifications to the impulse approximation
    contribution to the deuteron (left panel), $^3$He (middle panel)
    and $^4$He (right panel) $\sigma$-terms.  }
  \label{fig:fracSigma}
\end{figure}

\begin{figure}[!ht]
  \centering
  \includegraphics[width=0.45\textwidth]{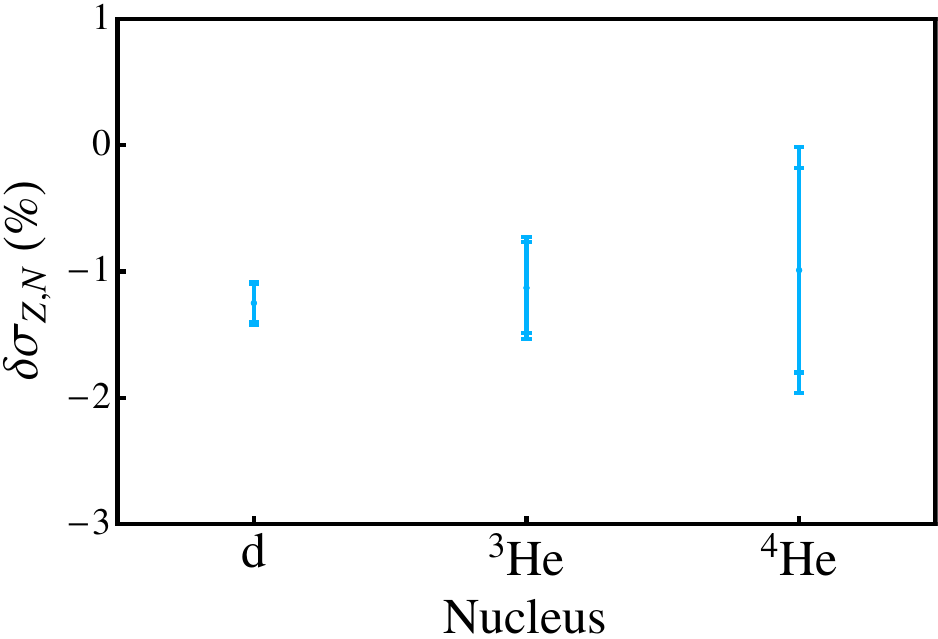}\
  \ \ \ \ \
  \includegraphics[width=0.45\textwidth]{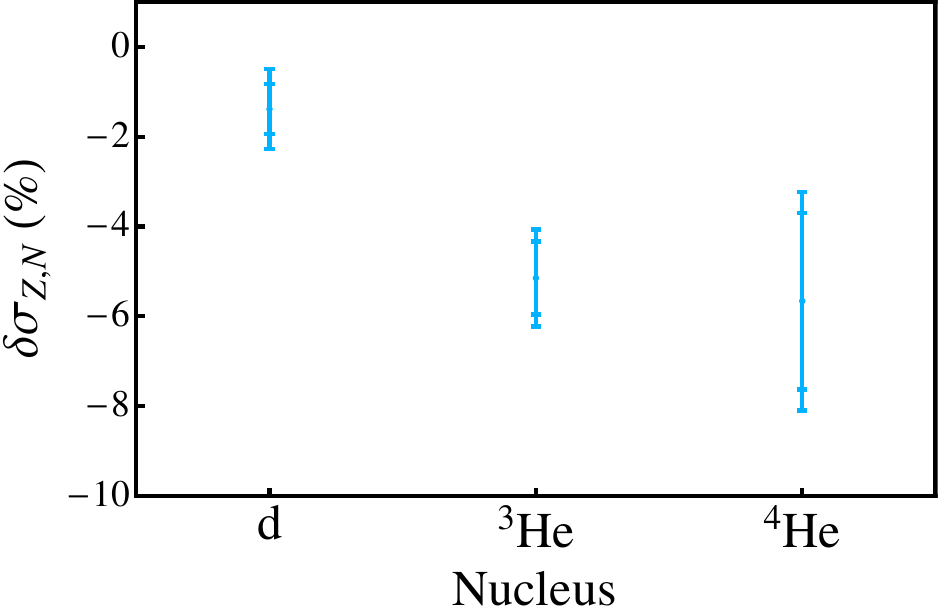}
  \caption{ Fractional contribution to the nuclear $\sigma$-terms from
    nuclear interactions for an average pion mass of $\langle
    m_\pi\rangle \sim 325~{\rm MeV}$ (left panel) and $\langle
    m_\pi\rangle \sim 658~{\rm MeV}$ (right panel).  }
  \label{fig:dsigVA}
\end{figure}

To estimate the derivative of the nuclear binding energies with
respect to the pion mass, as required in
Eq.~(\ref{eg:ZNsigdefisospinbinding}) to determine the nuclear
$\sigma$-terms, the finite differences between the binding energies at
adjacent pion masses are formed.  The extracted nuclear $\sigma$-terms
are given in Table~\ref{tab:SigmaTerms} and are shown in
Fig.~\ref{fig:SigmaB}; they take their values at the average pion mass
of the interval. This procedure assumes that there is negligible
curvature between adjacent pion masses; while purely linear behavior
is not expected, strong mass dependence is also not expected, as
determined from EFT and potential-model
phenomenology~\cite{Beane:2002vs,Epelbaum:2002gb,Beane:2002xf,Beane:2006mx,Soto:2011tb,Flambaum:2007mj,CarrilloSerrano:2012ja,Berengut:2013nh,Epelbaum:2012iu,Epelbaum:2013wla}. (See below for more discussion on this point.) To
determine the fractional contribution of the nuclear interactions to
the nuclear $\sigma$-terms, $\delta\sigma_{Z,N}$, the contribution
from the binding energy is divided by the contribution from
non-interacting nucleons, $A \sigma_N$. The single-nucleon mass is
known as a function of the pion mass from a broad range of lattice QCD
calculations, as summarized and analyzed recently in
Ref.~\cite{WalkerLoud:2008pj,Walker-Loud:2013yua}.  Empirically, it is
found that the results of lattice QCD calculations of the nucleon mass
are reasonably well reproduced by a linear dependence upon the mass of
the pion, $M_N = a_0 + a_1\
m_\pi$~\cite{WalkerLoud:2008pj,Walker-Loud:2013yua}, where
$a_0=802(13)~{\rm MeV}$ is the value in the chiral limit, and
$a_1=0.991(27)$, naively in conflict with expectations from HB$\chi$PT
which does not allow for a term linear in the pion mass.  Recent
analysis of lattice calculations near the physical pion mass indicate
that the next-to-leading order (NLO) expressions of HB$\chi$PT can
also be fit to the lattice results (for instance, see
Ref.~\cite{Bali:2012qs,Shanahan:2012wh,Alvarez-Ruso:2013fza,Durr:2011mp}),
and provide a nucleon $\sigma$-term at the physical pion mass that is
somewhat less than than estimate that follows from the empirical
linear relation.  Nevertheless, we conclude that the nucleon
$\sigma$-term is $\sigma_N = a_1 m_\pi/2$ to a good approximation in
the region of heavier pion masses where we extract the nuclear
$\sigma$-terms.\footnote{The empirical fit can be augmented to include
  higher-order polynomial terms without significantly altering the
  fitted $\sigma$-term. We also note that a very similar, though
  somewhat less precise, nucleon $\sigma$-term can be extracted using
  only the lattice calculations for which we have results for nuclei.}
\begin{table}
  \begin{center}
    \begin{minipage}[!ht]{16.5 cm}
      \caption{ Contributions to the nuclear $\sigma$-terms of the
        deuteron, $^3$He and $^4$He.  The binding energy
        contributions, $\sigma_{B_{Z,N}}$, are derived from the
        nuclear binding energies determined from lattice QCD
        calculations, shown in Table~\ref{tab:Masses}.  The quantity
        $\langle m_\pi \rangle$ is the average pion mass over the
        interval used to construct the finite-difference estimate of
        the nuclear $\sigma$-term.  The single-nucleon $\sigma$-term
        contribution, $A \sigma_N$, is taken from the approximate
        empirical relation $A \sigma_N = A a_1 m_\pi/2$, as defined in
        the text (with uncertainties determined from the covariance
        matrix of the two-parameter fit
        \protect\cite{Walker-Loud:2013yua}).  The first uncertainty of
        each quantity is statistical, the second is systematic and the
        third (where present) is the additional systematic associated
        with the relation between the pion mass and the light-quark
        mass.  }
      \label{tab:SigmaTerms}
    \end{minipage}
    \setlength{\tabcolsep}{1em} \resizebox{\linewidth}{!}{%
      \begin{tabular}{c|c|ccc}
        \hline
        $\langle m_\pi \rangle~({\rm MeV})$
        & Quantity
        & $d$  
        & ${^3 \rm He}$
        & ${^4 \rm He}$  \\
        \hline
        325
        & $A \sigma_N$ (MeV)
        & 322(9)(32)
        & 483(13)(48)
        & 644(17)(64)
        \\
        325
        & $\sigma_{B_{Z,N}}$ (MeV)
        & $-$4.08(48)(26)(41)
        & $-$5.5(1.8)(0.9)(0.6)
        & $-$6.5(5.3)(3.5)(0.7)
        \\
        325
        & $\delta\sigma_{Z,N}$
        & $-$0.0125(15)(08)
        & $-$0.0113(36)(18)
        & $-$0.0099(81)(54)
        \\
        \hline
        658
        & $A \sigma_N$ (MeV)
        & 652(18)(65)
        & 978(26)(98)
        & 1304(35)(130)
        \\
        658
        & $\sigma_{B_{Z,N}}$ (MeV)
        &  $-$9.1(3.7)(4.6)(0.9)
        & $-$50.8(8.0)(7.0)(5.1)
        & $-$75(26)(19)(8)
        \\
        658
        & $\delta\sigma_{Z,N}$
        & $-$0.0139(56)(70)
        & $-$0.0515(81)(71)
        & $-$0.057(20)(14)
        \\
        \hline
      \end{tabular}
    }
    \begin{minipage}[t]{16.5 cm}
      \vskip 0.0cm
      \noindent
    \end{minipage}
  \end{center}
\end{table}     
The fractional contribution of nuclear interactions to the nuclear
$\sigma$-terms of the deuteron, $^3$He and $^4$He are shown in
Fig.~\ref{fig:fracSigma}.  For each nucleus, the nuclear interactions
modify the $\sigma$-term by less than $10\%$ of the impulse
approximation contribution for both pion masses considered, and by
less than $2\%$ at the lighter pion mass, as can be seen in
Fig.~\ref{fig:dsigVA}.

\begin{figure}[!ht]
  \centering
  \includegraphics[width=0.59\textwidth]{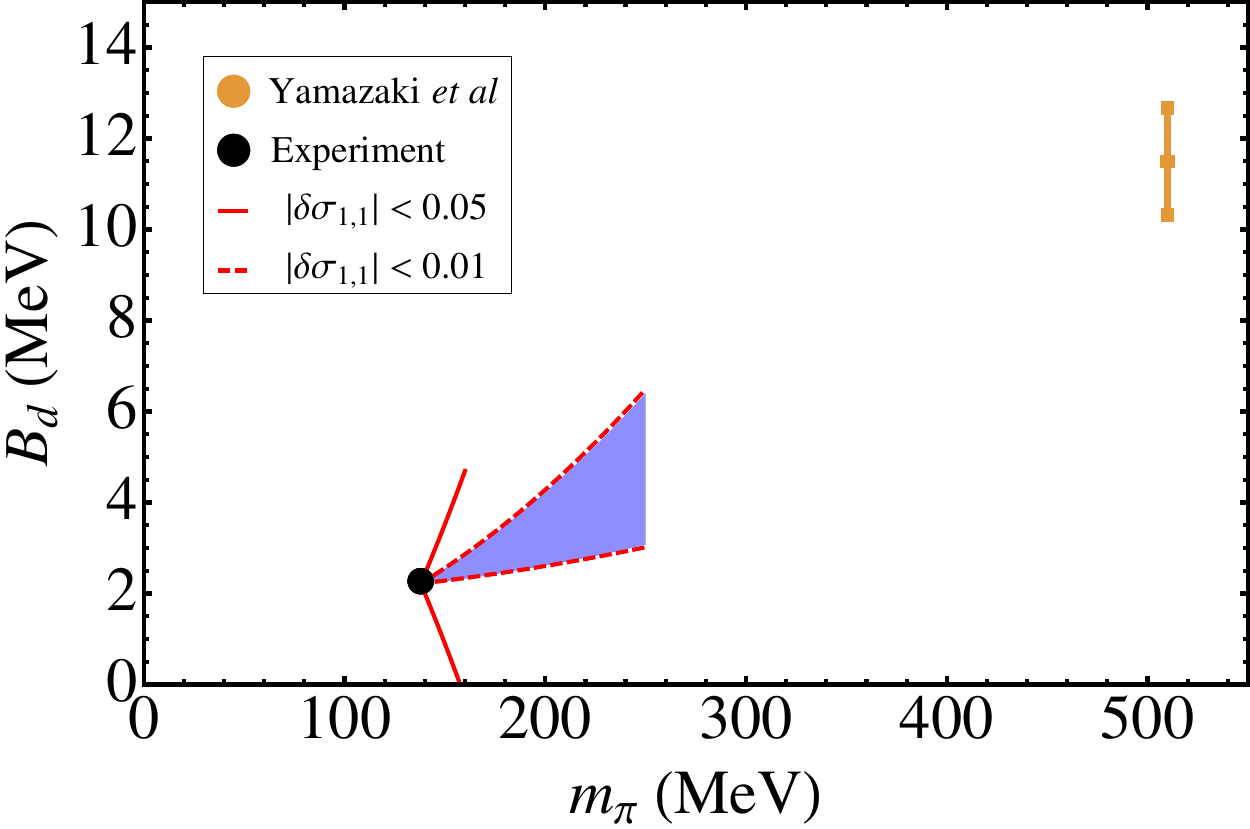}
  \caption{The deuteron binding energy from the $n_f=2+1$ lattice QCD
    calculation with the lightest pion mass~\protect\cite{Yamazaki:2012hi} and from
    experiment at the physical point. The band is from EFT together with naive dimensional analysis,
    from the enveloping curves (dashed lines) correspond to $|\delta\sigma_{1,1}|< 0.01$ and the
    boundary curves (solid lines) correspond to ($|\delta\sigma_{1,1}|< 0.05$).} 
  \label{fig:systerr}
\end{figure}

Given the heavy quark masses used in the existing lattice QCD
calculations of the binding energies of the light nuclei, it is worth
considering the systematic uncertainty due to extrapolation in the
light-quark masses. Using the EFT formalism for determining the
pion-mass dependence of the deuteron binding energy developed in
Refs.~\cite{Beane:2002vs,Beane:2002xf,Beane:2006mx},~\footnote{
In calculating the light-quark mass dependence of the deuteron binding energy,
Refs.~\cite{Beane:2002vs,Epelbaum:2002gb,Beane:2002xf}
use the quark-mass dependences of single hadron properties, such as $g_A$ and $f_\pi$,
determined from $\chi$PT.
As a result the estimated dependences of these quantities on the pion mass used in those works 
are much stronger than are now known from lattice QCD calculations.
Even so, within the uncertainties of the calculations, the deuteron binding energy could vary smoothly with the pion mass, as can be seen in 
Fig.~11 of Ref.~\cite{Epelbaum:2002gb} and Fig.~4 of Ref.~\cite{Beane:2002xf}.
In Ref.~\cite{Beane:2006mx}, the pion mass dependence of the single nucleon properties constrained by lattice QCD calculations were used to constrain behavior of the
scattering length in the $\siii$ channel, which was found to be consistent with a smooth variation of the deuteron binding energy, but with significant uncertainties.
} 
the coefficient
of the four-nucleon contact interaction (with an insertion of the
quark mass matrix) has been fit to pass through the central value of
the lattice QCD datum from Ref.~\cite{Yamazaki:2012hi}.  Fluctuations
in the value of this low-energy constant that are consistent with
naive dimensional analysis have been considered. The resulting band is
shown in Fig.~\ref{fig:systerr}. Calculating the deuteron sigma term
from the enveloping curves (dashed lines), we find that
$|\delta\sigma_{1,1}|< 0.01$ which suggests that the systematic
uncertainty in the binding energy due to extrapolation, translates to
at most a $1\%$ modification of the impulse approximation. We also
display boundary curves (solid lines) for a $5\%$ effect
($|\delta\sigma_{1,1}|< 0.05$).  One sees that the smooth, monotonic
behavior of the deuteron binding energy as a function of the pion mass
that is suggested by the lattice data (together with experiment) is
consistent with what is expected in the vicinity of the physical point
from EFT~\footnote{Similar considerations follow from an analogous EFT
  analysis of the singlet channel. While the dineutron is bound at
  heavy pion masses, a straight-line extrapolation is consistent with
  unbinding at the physical point, which suggests a very-shallow bound
  state, or a virtual bound state with large negative scattering
  length as seen in nature.}. Therefore the likelihood
of curvature significant enough to negate the conclusions found in
this paper is very small~\footnote{There may be significant curvature
  at pion masses below the physical point in the approach to the
  chiral limit, due to radiation pions~\cite{Soto:2011tb} or other
  effects that are non-analytic in the quark masses.}. A similar
analysis can be carried through for $^3$He and
$^4$He~\cite{Flambaum:2007mj} which also suggests smooth, monotonic
behavior in the quark masses in the vicinity of the physical point.

Deviations of the nuclear $\sigma$-terms from the sum of
single-nucleon contributions due to the nuclear forces are the same as
the deviations in the forward scalar-isoscalar WIMP-nucleus scattering
amplitudes from the sum of single nucleon amplitudes.  As such, the
results presented in this work are inconsistent with the conjectured
enhanced MEC contributions~\cite{Prezeau:2003sv} to WIMP-nucleus
scattering at the 10--60\% level.  Current lattice QCD calculations
indicate that deviations in the scalar-isoscalar WIMP-nucleus
interaction from interactions between the nucleons are at the percent
level, consistent with the typical size of MEC contributions.

\section{Discussion}
\label{sec:Conc}

\noindent 
Detecting and understanding dark matter is one of the great challenges
of our time. WIMPs endowed with weak-scale interactions arising from
relatively straightforward extensions to the Standard Model, are
natural candidates for dark matter.  A significant experimental effort
is ongoing to search for WIMPs, and one of the main techniques is to
search for nuclear recoils from WIMP-nucleus elastic collisions.
Calculations of the cross section for a WIMP-nucleus collision
typically involve determining the WIMP-hadron couplings in a hadronic
theory (with well-defined small expansion parameters) from the
fundamental WIMP-quark and WIMP-gluon interactions, and then taking
matrix elements of these hadronic-level operators between states in
the nucleus.  Experimental limits on WIMP interactions with ordinary
matter rely upon calculations of WIMP-nucleus cross sections based
upon the hadronic-level interactions.  It has been conjectured that
the scalar-isoscalar WIMP-nucleus cross section could be modified at
the 10--60\% level~\cite{Prezeau:2003sv} by meson-exchange currents
over naive expectations based upon the WIMP-nucleon interactions
alone.  Such terms would also scale differently with $Z$ and $N$ than
the impulse approximation. The uncertainty that such large MECs would
imply would lead to significant modifications to the reported limits
on this type of WIMP-nucleon interaction from experiment.  In this
work, we have combined the recent lattice QCD calculations of the
binding energies of the lightest few nuclei to provide a direct QCD
evaluation of nuclear $\sigma$-terms, albeit at unphysical quark
masses. These $\sigma$-terms are related to the amplitude for
WIMP-nucleus interactions in the scalar-isoscalar channel.  Deviations
from the single-nucleon contribution are found to be at the percent
level in light nuclei, inconsistent with the conjectured enhancement
but consistent with expectations based upon other nuclear observables,
such as electromagnetic moments and interactions.  Without enhanced
MEC contributions in light nuclei, it is likely that these effects are
also of their expected size in heavier nuclei and we expect that the
current nuclear calculations
\cite{Fitzpatrick:2012ix,Cirigliano:2012pq,Fitzpatrick:2012ib,Menendez:2012tm,Klos:2013rwa}
of WIMP-nucleus scattering cross sections are reliable. Additionally,
we note that the phenomenological nuclear $\sigma$-terms for various
nuclei estimated in
Refs.~\cite{Flambaum:2007mj,CarrilloSerrano:2012ja,Berengut:2013nh,Epelbaum:2012iu,Epelbaum:2013wla}
are also supported by our extractions. None of the lattice QCD
calculations of the light nuclei have been extrapolated to the
continuum and, as such, there are residual lattice-spacing
uncertainties in the results we have presented, however, these
systematic deviations from QCD are estimated to be small, and in
particular much smaller than other sources of uncertainty. Continuum
extrapolated calculations will be performed in the near future.

The conclusions of this work were obtained by assuming  that the nuclear binding energies vary smoothly 
with the light-quark masses.
In the two-nucleon sector, a comparison between this dependence constrained by EFT 
for light pions and the results of Lattice QCD at heavier pion masses, outside the range of applicability of the EFT,
is consistent with this assumption.
While, there is no indication that this assumption should not hold for the nuclei considered in this work, 
it does introduce a systematic uncertainty into our results that, 
while we have attempted to quantify its impact,
remains to be eliminated by future Lattice QCD calculations. 
Our use of naive dimensional analysis to determine the range of coefficients in the EFT provides only an estimate 
of their values, and the true values could be somewhat larger or somewhat smaller than predicted by 
NDA - there are well-known examples of both in nature.  
Therefore, it is possible that the estimate of the 
systematic uncertainty that we have assigned to the extrapolation is somewhat larger or somewhat smaller 
than its actual value.
Better quantifying the extrapolation in light-quark masses through improved 
Lattice QCD calculations over a range of light-quark masses,
and better understanding the theoretical form of the extrapolation
is a focus of our ongoing program, and also that of others.

To help understand the disagreement of our results with the
predictions of Ref.~\cite{Prezeau:2003sv}, we consider the problem
using KSW power-counting in nuclear EFTs in the
$\si$-channel~\cite{Kaplan:1996xu,Kaplan:1998tg,Kaplan:1998we}.  While
the $\si$ channel does not exhibit a bound state at the physical
point, a bound-state develops as the pion mass is
increased~\cite{Beane:2009py,Beane:2011iw,Yamazaki:2012hi,Beane:2012vq,Yamazaki:2012fn}.
More importantly, KSW power-counting permits an analytic analysis of
the scattering amplitude and bound state properties.  The expansion of
the dineutron $\sigma$-term is analogous to the expansion of the
electromagnetic form factors and moments of the deuteron with KSW
power counting~\cite{Kaplan:1998sz}, in which it is straightforward to
show that the four-nucleon operators, and more generally the NLO
contributions such as the MEC contributions, are suppressed by factors
of the deuteron binding momentum divided by the EFT cutoff scale
compared with the LO single-nucleon contribution.  In electroweak
processes, deviations from the single-nucleon contribution are at the
percent level, and it is reasonable to expect similar modifications
for the scalar current. This suggests that the arguments of
Ref.~\cite{Prezeau:2003sv} regarding the importance of the MECs fail
explicitly in this channel. Unfortunately, the chiral expansion of the
nuclear potential \`a la Weinberg is formally
inconsistent~\cite{Kaplan:1996xu}, and KSW power counting is invalid
in the $\siii$--$\diii$ coupled-channels.  Therefore, an analysis of
the behavior of observables based upon the nuclear potential alone
cannot be reliably performed. It was such an analysis that was
performed in Ref.~\cite{Prezeau:2003sv} in order to arrive at the
conjecture of parametrically enhanced MECs for the scalar-isoscalar
WIMP-nucleus interactions.  Current lattice results show that this
analysis does indeed fail in the cases under consideration.

It is clear that a systematic study of the light-quark mass dependence
of the lightest nuclei is a valuable program to pursue, not only for
intellectual reasons, but also in order to reduce the uncertainty in
the response of nuclei to probes other than those provided by the
electroweak interactions.  An important application of these
calculations is the refinement and complete quantification of the
nuclear physics component of WIMP-nucleus interactions that are
crucial for interpreting the results from many experiments searching
for dark matter.  We anticipate that lattice QCD calculations of a
wider range of WIMP-nucleus interactions in the light nuclei will
become possible in the near future as larger computational resources
become available.  While the scalar-isoscalar interactions in the
forward direction arising from fundamental quark interactions with
WIMPs can be accessed through the light-quark mass dependence of the
masses of the nuclei, the matrix elements of other operator structures
in light nuclei will require forming three-point correlation
functions, or the use of background-field techniques.  The results of
these calculations will refine the structure of the effective
interactions and will enable the determination of WIMP-nucleus
interactions with reduced uncertainties.

\acknowledgments We would like to thank W.~Haxton, T.~C.~Luu,
M.~McCullough, S.~Meinel, U.-G.~Mei\ss ner, K. Orginos, A. Schwenk, and
A. Walker-Loud for useful discussions.  SRB was partially supported by
NSF continuing grant PHY1206498.  In addition, SRB gratefully
acknowledges the hospitality of the Helmholtz-Institut f\"ur Strahlen-
und Kernphysik at the University of Bonn, and the Mercator programme
of the Deutsche Forschungsgemeinschaft.  WD was supported by the
U.S. Department of Energy through Outstanding Junior Investigator
Award DE-SC000-1784.  H-WL and MJS were supported in part by the DOE
grant DE-FG03-97ER4014.  MJS thanks the Alexander von Humboldt
foundation for the award that enabled his visit to the University of
Bonn, and the kind hospitality of U.-G.~Mei\ss ner and the
Helmholtz-Institut f\"ur Strahlen- und Kernphysik at the University of
Bonn.


\end{document}